\documentclass[12pt]{article}
\begin{document}

\newcommand{\Mmn}{M_{mn}}
\newcommand{\Mkl}{M_{kl}}
\newcommand{\Sumn}{\sum\limits_{mn}}
\newcommand{\Sukl}{\sum\limits_{kl}}

\newcommand{\Omn}{\Omega_{mn}}
\newcommand{\Okl}{\Omega_{kl}}

\newcommand{\TE}{\mathrm{TE}}
\newcommand{\TM}{\mathrm{TM}}

\newcommand{\sigep}{4\pi\sigma}
\newcommand{\ez}{\bar{\varepsilon}}
\newcommand{\ve}{\varepsilon}
\newcommand{\tG}{\tilde{\Gamma}}
\newcommand{\Am}{A_{\mu}}
\newcommand{\Amsq}{A_{\mu}^{(2)}}
\newcommand{\Amqu}{A_{\mu}^{(4)}}

\newcommand{\Li}{\mathrm{Li}}

\newcommand{\la}{\label}
\newcommand{\bea}{\begin{eqnarray}}
\newcommand{\eea}{\end{eqnarray}}

\newcommand{\be}{\begin{equation}}
\newcommand{\ee}{\end{equation}}






\begin{center}
\large{{\bf Van der Waals interactions: Evaluations by use of a statistical mechanical method}}
 
\vspace{.5cm}
Johan S. H{\o}ye\footnote{johan.hoye@ntnu.no}

Department of Physics, Norwegian University of Science and Technology, N-7491 Trondheim, Norway

\today

\end{center}
\ \\

\begin{abstract}

In this work the induced van der Waals interaction between a pair of neutral atoms or molecules is considered by use of a statistical mechanical method. Commonly this interaction is obtained by standard quantum mechanical perturbation theory to second order. However, the latter is restricted to electrostatic interactions between charges and dipole moments. So with radiating dipole-dipole interaction where retardation effects are important for large separations of the particles, other methods are needed, and the resulting induced interaction is the Casimir-Polder interaction usually obtained by field theory. It can also be evaluated, however, by a statistical mechanical method that utilizes the path integral representation. 
We here show explicitly by use of the statistical mechanical method the equivalence of the Casimir-Polder and van der Waals interactions to leading order for short separations where retardation effects can be neglected. Physically this is well known, but in our opinion the mathematics of this transition process is not so obvious. The evaluations needed mean a transform of the statistical mechanical free energy expression to a form that can be identified with second order perturbation theory. In recent works [H{\o}ye 2010] the Casimir-Polder or Casimir energy has been added as a correction to calculations of systems like the electron clouds of molecules.
 The equivalence to van der Waals interactions to leading order indicates that the added Casimir energy will improve the accuracy of calculated molecular energies. We here also give numerical estimates of this energy including analysis and estimates for the uniform electron gas.

\end{abstract}
{\small
Key words: Polarizability, path integral, Casimir interaction, perturbation theory.\\
PACS numbers: 03.65.-w, 05.20.-y, 05.30.-d, 34.20.Gj}


\newpage

\section{Introduction}
\label{intro}

The well known Casimir effect is commonly regarded to be due to fluctuations of the quantized electromagnetic field in vacuum and has been widely studied in this respect. However, H{\o}ye and Brevik considered this in a different way by regarding the problem as a statistical mechanical one of interacting fluctuating dipole moments of polarizable particles. In this way the Casimir-Polder force  \cite{casimir48} between a pair of polarizable point particles was recovered \cite{brevikhoye88}. To do so the path integral formulation of quantized particle systems was utilized \cite{feynman53}. With the path integral the quantum mechanical problem at thermal equilibrium becomes equivalent to a classical polymer problem in four dimensions where imaginary time is the fourth dimension.  The role of the electromagnetic field is then to mediate the pair interaction between polarizable particles. Later this type of evaluation was generalized to a pair of parallell plates, and the well known Lifshitz result was recovered \cite{hoyebrevik98}. Similar evaluations were performed for other situations \cite{hoyebrevik00, hoyebrevik01}.
Earlier this method was fruitfully utilized for a polarizable fluid \cite{hoyestell81}.

The statistical mechanical approach opened new perspectives for evaluations of the Casimir force. Instead of focusing upon the quantization of the electromagnetic field itself one can regard the problem as one of polarizable particles interacting via the electromagnetic field. It was found that these two viewpoints are equivalent \cite{brevikhoye88, hoyebrevik98, hoyebrevik01, hoyebrevikaarsethmilton03}.

The source of the Casimir free energy is the radiating electromagnetic interaction. For large separations of a pair of atoms or molecules retardation effects are important. Thus methods of field theory have been widely used to obtain the resulting Casimir force or energy for various situations \cite{spruch93}. Such situations are a pair of atoms, an atom and a wall, or two walls separated by some distance. Further the Casimir energy influences the shift of energy levels of atoms, and it is closely related to the Lamb shift as they both are consequences of the vacuum fluctuations of the electromagnetic field \cite{spruch93}.

For short separations, however, retardation effects can be neglected, and on physical grounds it is well known that the Casimir interaction becomes the van der Waals interaction in this case. The usual way to compute the latter is to apply second order perturbation theory based upon the Schr\"{o}dinger equation.

Although the equivalence or the transition from Casimir to van der Waals interaction is well known from the physical point of view, it is not obvious or trivial from the mathematical side as the methods used are very different. To obtain the Casimir interaction summation over imaginary Matsubara frequences are performed and the frequency dependent polarizability is needed. On the other side the van der Waals interaction can be expressed directly in terms of the matrix elements and eigenvalues of second order perturbation theory. Thus in this work we will show explicitly the equivalence of the Casimir and van der Waals interactions in the case of static interactions. To do so the statistical mechanical free energy expression for the Casimir energy to leading order is transformed such that it can be identified with second order perturbation theory.
The equivalence obtained coincides with the interpretation that the Casimir force can be related to fluctuating dipole moments. The statistical mechanical method introduced in Ref.~\cite{brevikhoye88} is based upon this interpretation realizing that the electromagnetic field can be fully replaced by pair interactions between dipolar moments \cite{hoyebrevik98,hoyebrevik00,hoyebrevik01,hoyebrevikaarsethmilton03}.

The mathematical equivalence of the Casimir and van der Waals interactions, which we will show below, gives a close and direct connection between the Casimir theory and standard quantum mechanical perturbation theory. This means that the Casimir interaction can be interpreted as a perturbing correction upon quantized many-body systems of interacting particles. Thus it is of interest to take this energy into account for  a more general situation of such systems like electrons in molecules. This direct and close connection seems not to have been noticed previously. Radiation effects, if taken into account, will be minor corrections. Recently work in this direction was initiated by Buenzli and Martin \cite{buenzlimartin08} and by Jancovici and \v{S}amaj \cite{jancovicisamaj09}. 

Earlier the path integral formulation of fermion and boson fluids was considered by Chandler and Wolynes \cite{chandlerwolynes81} and by H{\o}ye and Stell \cite{hoyestell94}. But the latter investigations were restricted to uniform density and instantaneous interactions. Present standard methods to evaluate many-electron systems utilize Hartree-Fock (HF) and density functional theory (DFT)\cite{kohnsham65}. By these latter methods exchange effects are taken into account, and  approximations for local correlations are made. In separate works we have recently reconsidered this problem by adding the corresponding Casimir interaction to ab initio calculations of interacting many-body systems like electrons of molecules \cite{hoye10, hoye10a}. In Ref.~\cite{hoye10a} we also were able to include the radiating electromagnetic field, and it was found that an ionic fluid of electrons bound to atomic nuclei can be identified with a dielectric one.



With the Casimir energy included non-local correlations between electrons are accounted for. These correlations correspond to the induced ones between the fluctuating dipole moments (or polarizability) of the pair of point particles considered in the this work. To estimate the influence of these correlations we will make crude estimates for atoms at close contact, and further make an analytic approximation with numerical estimates for the uniform electron gas.


To compare the statistical mechanical method with standard second order perturbation theory some obvious results of the latter are needed and are therefore established below.
Thus in Sec.~\ref{sec2} we write down the well known expression for the static polarizability due to a particle or electron bond by a potential. This is expressed in a standard way by means of energy eigenvalues and matrix elements between eigenstates. 

In Sec.~\ref{sec3} the appropriate pair correlation function in imaginary time needed for the statistical mechanical method is obtained. To do so the correlation function in imaginary time is established in view of the path integral, and Kubo relations are used. With the simplified dipolar type interaction used in Sec.~\ref{sec4} one finds that its Fourier transform in imaginary time is the well-known dynamic polarizability.

In  Sec.~\ref{sec4} the statistical mechanical method is used to obtain the Casimir interaction between a pair of particles. This is the result we want to compare with standard quantum mechanical perturbation theory to second order for static interactions. To do so the leading term in the Casimir interaction is considered and explicit summations with respect to the Matsubara frequencies are performed. Then the expressions obtained for the dynamical polarizability are used, and we show that the Casimir energy can be expressed in terms of the usual matrix elements of quantum mechanics. The result depends upon temperature in general. 

In Sec.~\ref{sec5} the statistical mechanical result obtained in Sec.~\ref{sec4} is verified to be equivalent to standard quantum mechanical perturbation theory to second order.

In Sec.~\ref{sec6} the situation is generalized in a straightforward way to more particles or electrons bond to a potential. Then properties of fermions (or bosons) have to be taken into account to obtain the appropriate correlation function in imaginary time. Again this is found to be the corresponding dynamical polarizability. 

In Sec.~\ref{sec7} the statistical mechanical approach is again identified with second order perturbation theory for the more general situation with several electrons, which is most relevant for realistic many-body systems.  

In Sec.~\ref{sec8} numerical estimates of the Casimir energy for neighboring pairs of atoms in molecules due to induced correlations are made. Then a pair of polarizable particles at close contact, but otherwise regarded as point particles, are considered.

In Sec.~\ref{sec9} the homogeneous  electron gas is considered and an analytic estimate of the correction to ab initio calculations are made. An explicit approximation that covers all electron densities is obtained. Especially for low density it is found that the ground state energy of plasma oscillations is the main contribution besides the binding energy. One may speculate whether this has relation to high-$T_c$ super-conductivity that is considered to be connected to properties of electrons at low density.

In Sec.~\ref{sec10} numerical estimates of the analytic result obtained in Sec.~\ref{sec9} are made. One estimate is for the density of conduction electrons in a metal. In this connection it is noted that  the Casimir energy can be associated with the random phase approximation (RPA) for quantized many-body systems. Work has been performed on the RPA to improve it further in view of best available "exact" results. So the analytic approximation of Sec.~\ref{sec9} is compared with one such result for high electron density. Further the statistical mechanical viewpoint indicates new ways to improve the RPA. This will be discussed shortly.

\section{Static polarizability}
\label{sec2}

The statistical mechanical method requires the appropriate correlation function of a reference system as input. In the present case with pair interactions of dipolar type this will be the dynamical polarizability as derived in the next section. First we will recover the simple result for the static polarizability. So
consider a particle in an external potential like an electron bound to an atomic nucleus. It is perturbed by an electric field ${\cal E}_x$ along the $x$-axis. This contributes
\be
H^{'}=-qx{\cal E}_x
\la{1}
\ee
to the Hamiltonian where $q$ is charge and $x$ is position. The applied field ${\cal E}_x$ produces a net dipole moment $P_x$ in the same direction provided the unperturbed system possesses spherical symmetry. By standard perturbation theory one easily obtains the static polarizability $\alpha$ either via the average position $\langle x\rangle$ as $P_x=q\langle x\rangle=\alpha{\cal E}_x$ or via the perturbed energy to second order given by $E_2=-\frac{1}{2}P_x {\cal E}_x$. In either case one finds \cite{london30}
\be
\alpha=2\sum\limits_m\frac{M_{m0}}{\Omega_{m0}}
\la{2}
\ee
provided the particle is located in the ground state as is the case for low temperatures, $T\rightarrow 0$. Here we have introduced the abbreviations
\bea
\nonumber
\Mmn&=&q^2\langle\phi_m|x|\phi_n\rangle\langle\phi_n|x|\phi_m\rangle\\
\Omn&=&E_m-E_n
\la{3}
\eea
where the $\phi_m=\phi_m (x)$ are the unperturbed wavefunctions and $E_m$ are the corresponding eigenvalues for the energies. The $\phi_0$ and $E_0$ are ground state quantities. It can be noted that in this work all states are considered to be bound states of the potential to avoid possible problems connected to unbound states with continuous spectrum.

\section{Dynamic polarizability}
\label{sec3}

With the path integral representation the correlation function of interest can be obtained from the one particle correlation function in imaginary time $\lambda$. This function is given by \cite{hoye10,hoye10a}
\be
G(\lambda,x_1, x_2)=\frac{1}{Z}\Sumn F_{mn}(\lambda,x_1,x_2)
\la{4}
\ee
where
\bea
\nonumber
F_{mn}(\lambda,x_1,x_2)&&= \phi_m (x_1)e^{-(\beta-\lambda) E_m}\phi_m^* (x_2)\phi_n (x_2)e^{-\lambda E_n}\phi_n^* (x_1)\\ 
\quad \mbox{and} \quad Z&&=\sum\limits_k e^{-\beta  E_k} 
\la{4b}
\eea
recognizing that in the path integral picture the quantized particle represents a "classical" polymer in the imaginary time direction. In this direction it has an extension $\beta=1/(k_B T)$ where $T$ is temperature and $k_B$ is Boltzmann's constant. The relation between $\lambda$ and real time $t$ is $t=-i\hbar \lambda$.
(For simplicity one dimensional notation is used here and below. By a mistake the $\lambda$ was defined with opposite sign in Ref.~\cite{brevikhoye88}, but this did not influence previous applications where (apart from a constant) the operators $A$ and $B$ of Eq.~(\ref{7}) below were equal. The mistake was corrected in Appendix of Ref.~\cite{hoye10a}.)

The dipole moment $P_x$ and thus $\alpha$ is now found from the appropriate response function via the Kubo relations \cite{brevikhoye88,hoye10a,kubo58}. Then we need the average
\bea
\nonumber
g(\lambda)&=&q^2\langle x_1 x_2\rangle=q^2\int x_1 x_2 G(\lambda, x_1, x_2)\, dx_1\,dx_2\\
&=&\Sumn p_m \Mmn e^{\lambda\Omn}
\la{5}
\eea
where
\be
p_m=e^{-\beta E_m}/Z \quad (\sum\limits_m p_m=1).
\la{5b}
\ee
The $p_m$ is the probability for the particle to occupy energy level $m$ at thermal equlibrium. Further the Fourier transform is needed
\bea
\nonumber
\tilde{g}(K)&=&\int\limits_0^\beta g(\lambda)e^{iK\lambda}\,d\lambda=\Sumn(p_n-p_m)\Mmn\frac{1}{\Omn+iK}\\
&=&\Sumn(p_n-p_m)\frac{\Mmn\Omn}{\Omn^2+K^2}
\la{6}
\eea
where the symmetry properties $\Mmn=M_{nm}$ and $\Omn=-\Omega_{nm}$ have been used.
The $K=2\pi n/\beta$ with $n$ integer are the Matsubara frequencies.

The resulting dipole moment  $P_x=q\langle x\rangle$ is now related to the applied (time varying) electric field ${\cal E}_x$ via the response function \cite{brevikhoye88,hoye10a,kubo58}
\be
\varphi_{BA}(t)=\frac{1}{i\hbar}{\rm Tr}(\rho[A,B(t)])
\la{7}
\ee
where $\rho$ is the canonical equilibrium density matrix $\rho=e^{-\beta H}/Z$ ($Z=Tr(e^{-\beta H})$), and the $B(t)$ in the commutator is the time-dependent operator $B(t)=e^{\lambda H/\hbar}Be^{-\lambda H/\hbar}$ where $\lambda=it/\hbar$. For a time-dependent perturbation $H^{'}=-AF(t)$ the change in the quantity $B$ is then given by
\be
\langle \Delta B(t)\rangle=\int\limits_{-\infty}^t \varphi_{BA}(t-t^{'})\,F(t^{'})\,dt^{'}.
\la{8}
\ee
As noted above the $\lambda$ by a mistake was defined with opposite sign in Ref.~\cite{brevikhoye88}, but it was corrected in Ref.~\cite{hoye10a}. By this correction the Schr{\"o}dinger equation in imaginary time transforms to an equation that describes diffusion for increasing $\lambda$. With this the correlation function in imaginary time and response function in real time are
\bea
\nonumber
g(\lambda)&=&
\rm{Tr}(\rho B(t)A) \\
\la{9} \varphi_{BA}(t)&=&\frac{1}{i\hbar}\left[g(\beta+\lambda)-g(\lambda)\right]
\eea

As found in Appendix B in Ref.~\cite{brevikhoye88} the Fourier transform of the correlation function in imaginary time equals the Fourier transform of the response function in real time, i.e.~
\be
\tilde\varphi_{BA}(\omega)=\tilde g (K)
\la{10}
\ee
in the common region where Im$(\omega)<0$ and $|$Im$(K)|<C$ $(C>0)$ with $K=  i\hbar\omega$.

In the present case with perturbation (\ref{1}) and polarization $P_x=q\langle x\rangle=\langle \Delta B\rangle$ we have the operators $A=qx=qx(0)$ and $B(t)=qx(t)$ by which the correlation function will be the one given by Eq.~(\ref{5}) with Fourier transform (\ref{6}). With $\tilde P_x(\omega)=\alpha(K) {\cal E}_x(\omega)$ ($K=i\hbar\omega$) it thus follows that the frequency dependent polarizability is given by
\be
\alpha=\alpha(K)=\tilde g(K).
\la{11}
\ee
For $K=0$ and zero temperature the static result (\ref{2}) is recovered.

\section{Induced interaction between a pair of particles}
\label{sec4}

Consider a pair of particles located at different positions in space each with polarizability given by (\ref{11}). Then assume there is a mutual perturbing interaction between these particles given by
\be
\Phi=\psi q^2 x_a x_b
\la{12}
\ee
where $\psi=\psi(r)$ is a quantity that depends upon spatial separation while $x_a$ and $x_b$ are internal coordinates of the two separate binding potentials. Interaction (\ref{12}) is similar to the dipole-dipole interaction where $qx_a$ and $qx_b$ correspond to dipole moments restricted to the $x$-direction. We will use the one dimensional interaction (\ref{12}) instead of the full dipolar interaction to simplify the derivations below a bit. The purpose is to transform the statistical mechanical free energy expression to a form that can be identified with second order perturbation theory.

The free energy change $\Delta F$ due to the interaction can be found from \cite{brevikhoye88}
\be
\Delta F=\frac{1}{2\beta}\sum\limits_K \ln(1-(\alpha\psi)^2)\approx -\frac{1}{2\beta}\sum\limits_K(\alpha\psi)^2
\la{13}
\ee
to second order in $\alpha=\alpha(K)$. For $K=0$ expression (\ref{13}) is just the classical expression for the free energy change due to a pair of classical harmonic oscillators that interact. This is Eq.~(3.4) of Ref.~\cite{brevikhoye88}. In the quantum mechanical case one has to sum over the Matsubara frequencies $K=2\pi n/\beta$ ($n$ is integer) as follows from the path integral. More generally the interaction can be time-dependent like the radiating dipole interaction. Then the Fourier transform $\tilde \psi (K)$ with $K=i\hbar\omega$ is needed. However, to enable comparison with standard quantum mechanical perturbation theory the statistical mechanical expression should be restricted to the static case, i.e.~$\psi=\tilde \psi(0)$ for all $K$.

It can be noted that the term to second order in $\alpha$ in expression (\ref{13}) is consistent with the original derivation of the Casimir-Polder force \cite{casimir48} when modified such that the $\psi$ is replaced with the radiating dipolar interaction that depends upon orientation of dipolar moments and frequency too, as done in Ref.~\cite{brevikhoye88}. Casimir and Polder used the methods of quantum electrodynamics to obtain this force. Wennerstr\"{o}m et al.~made a closer study of the temperature dependence of the Casimir interaction where both $\alpha$ and $\psi$ depend upon $K$ \cite{wennerstrom99}. It can be noted that their expression (5) (except for the $n=0$ term that is taken out) for this interaction between a pair of particles is the same as expression (5.15) obtained by the statistical mechanical method in Ref.~\cite{brevikhoye88}. (A factor 2 is missing in the denominator of Eq.~(5.15) of the reference, but is present in its Eq.~(5.14).) This is seen when the $\tau$ and $\alpha_K$ of Ref.~\cite{brevikhoye88} are identified with the $\pi xn$ and $\alpha(0)/(1+(An)^2)$ of Ref.~\cite{wennerstrom99} where $A=2\pi k_B T/(\hbar\omega_0)$.) 

In the present work we want to verify that the Casimir interaction for the specific case of static interactions, i.~e.~when $\psi$ does not depend upon $K$, is identical to the van der Waals interaction that follows from standard quantum mechanical perturbation theory to second order. Then expression (\ref{13}) must be transformed to a form that is recognizable and can be identified with the latter. This is done by explicit summations in (\ref{13}) with respect to the Matsubara frequencies $K=2\pi n/\beta$. We need the sum ($K=2\pi n/\beta$)
\be
\frac{1}{\beta}\sum\limits_K \frac{\Omega}{\Omega^2+K^2}=\frac{\cosh(\beta\Omega/2)}{2\sinh(\beta\Omega/2)}.
\la{14}
\ee
It follows from the backward transform of expression (\ref{6}) for $\lambda=0$ noting that the result should be the average of $(g(0)+g(\beta))/2$ since implicitly the $g(\lambda)$ is assumed periodic. With this we obtain
\bea
\nonumber
&&\frac{1}{\beta}\sum\limits_K \frac{\Omega_1}{\Omega_1^2+K^2}\frac{\Omega_2}{\Omega_2^2+K^2}= \frac{\Omega_1 \Omega_2}{\Omega_2^2-\Omega_1^2} \frac{1}{\beta}\sum\limits_K \left[\frac{1}{\Omega_1^2+K^2}-\frac{1}{\Omega_2^2+K^2}\right]\\
\nonumber
&&=\frac{1}{2(\Omega_2^2-\Omega_1^2)}\left[\Omega_2\frac{\cosh(\beta\Omega_1/2)}{\sinh(\beta\Omega_1/2)}-\Omega_1\frac{\cosh(\beta\Omega_2/2)}{\sinh(\beta\Omega_2/2)}\right]\\
\la{16}
&&=\frac{1}{4\sinh(\beta\Omega_1/2)\sinh(\beta\Omega_2/2)}\\
\nonumber
&&\times\left[\frac{\sinh(\beta(\Omega_1+\Omega_2)/2)}{\Omega_1+\Omega_2}+\frac{\sinh(\beta(\Omega_1-\Omega_2)/2)}{\Omega_1-\Omega_2} \right].
\eea
In the limit $\beta\rightarrow\infty$ this result simplifies to $1/[2(|\Omega_1|+|\Omega_2|)]$.

By use of expressions (\ref{6}) and (\ref{16}) we now find for the free energy contribution (\ref{13})
\bea
\nonumber
&&\Delta F=-\frac{\psi^2}{2\beta}\sum_K [g(K)]^2
=-\frac{\psi^2}{2} \Sumn\Sukl \Mmn\Mkl(p_m p_n p_k p_l)^{1/2}\\
\nonumber &&\times\left[\frac{\sinh(\beta(\Omn+\Okl)/2)}{\Omn+\Okl}+\frac{\sinh(\beta(\Omn-\Okl)/2)}{\Omn-\Okl}\right]\\
\nonumber
&&=-\frac{\psi^2}{4} \Sumn\Sukl \Mmn\Mkl\left[\frac{p_n p_l-p_m p_k}{\Omn+\Okl}+\frac{p_n p_k-p_m p_l}{\Omn-\Okl}\right]\\
&&=-\psi^2 \Sumn\Sukl p_n p_l \frac{\Mmn\Mkl}{\Omn+\Okl}.
\la{17}
\eea
Here $p_n-p_m= 2(p_n p_m)^{1/2}\sinh(\beta\Omn/2)$ has been used ($p_n=e^{-\beta E_n}/Z$).
The last equality in Eq.~(\ref{17}) follows by use of the symmetries $\Mmn=M_{nm}$ and $\Omn=-\Omega_{nm}=E_m-E_n$ by summations. 

\section{Second order perturbation theory}
\label{sec5}

For comparison the expressions of standard quantum mechanical perturbation theory are needed. For two separate particles with interaction (\ref{12}) the perturbed energy of the eigenstates $n$ and $l$ is
\be
\Delta E_{nl}=-(q^2 \psi)^2\sum_{mk}\frac{|\langle mk|x_a x_b|nl\rangle|^2}{E_m+E_k-(E_n+E_l)}=-\psi^2\sum_{mk}\frac{\Mmn\Mkl}{\Omn+\Okl}
\la{18}
\ee
to second order. Here $\langle mk|x_a x_b|nl\rangle=\langle m|x_a|n\rangle\langle k|x_b|l\rangle$ is used. (The contribution to first order with $m=n$ and $k=l$ i assumed zero.)
For $T=0$ it is easily seen that the free energy (\ref{17}) coincides with expression (\ref{18}) for $\Delta E_{00}$  as then the system is in its ground state with $n=l=0$.

For finite $\beta$ one should consider the resulting partition function
\bea
\nonumber
&&Z_r=\sum_{nl} e^{-\beta(E_n+E_l+\Delta E_{nl})}=\sum_{nl} e^{-\beta(E_n+E_l)}(1-\beta \Delta E_{nl}+\dots)\\
&&=Z^2\left(1-\frac{\beta}{Z^2}\sum _{nl}e^{-\beta(E_n+E_l)} \Delta E_{nl}+\dots\right).
\la{19}
\eea
With free energy $F=-\ln Z_r/\beta$ the perturbed part of it will be
\be
\Delta F=\frac{1}{Z^2}\sum_{nl}e^{-\beta(E_n+E_l)}\Delta E_{nl}+\cdots.
\la{20}
\ee
where expression (\ref{18}) is to be inserted. One sees that the statistical mechanical result (\ref{17}) coincides with this equation. This verifies the mathematical equivalence  of the two methods for static interactions.

\section{Polarizability with several particles}
\label{sec6}

With several particles in a common potential, e.g.~electrons in an atom, one has to take into account the symmetry properties of fermions (or bosons). However, the eigenstates can again be represented (approximately at least) by one-particle wave functions. But the correlation function (\ref{4}) is modified into
\be
G(\lambda, x_1, x_2)=\zeta\Sumn f_n f_m F_{mn}(\lambda, x_1,x_2).
\la{21}
\ee
Here
\be
f_n=\frac{1}{1+\zeta e^{-\beta E_n}} \quad \mbox{and} \quad\zeta=e^{\beta\mu}
\la{22}
\ee
where $\mu$ is the chemical potential. (For bosons $f_n=1/(1-\zeta e^{-\beta E_n})$.) The $f_n$ factors include all sums of path integral polymers that are tied together into coils. 

With (\ref{21}) it immediately follows that the expression for the polarizability as given by expression (\ref{6}) can be kept if $p_n$ for fermions is redefined to be
\be
p_n=f_n \zeta e^{-\beta E_n}=\frac{\zeta e^{-\beta E_n}}{1+\zeta e^{-\beta E_n}}
\la{25}
\ee
which is the probability for level $n$ to be occupied by a particle. (Thus definition (\ref{25}) for $p_n$ replaces the one by Eq.~(\ref{5b}).) With this we again have
\be
\alpha(K)=\tilde g(K)=\Sumn (p_n-p_m)\frac{\Mmn\Omn}{\Omn^2+K^2}.
\la{23}
\ee
noting that 
$\zeta f_m f_n (e^{-\beta E_n}-e^{-\beta E_m})=p_n -p_m$.

It may be noted that expression (\ref{23}) can be interpreted as a sum of independent contributions from each of the particles (or electrons) present. This reflects the one-particle nature of the wave functions used. This is seen by again using $\Mmn=M_{nm}$ and $\Omn=-\Omega_{nm}$ by which the $-p_m$ in Eq.~(\ref{23}) can be replaced with $p_n$ to obtain
\be
\alpha (K)=2\Sumn p_n \frac{\Mmn\Omn}{\Omn^2+K^2}.
\la{26}
\ee
Further the average number of particles that contribute to $\alpha(K)$ is $N=\sum_n p_n$. Thus expression (\ref{26}) shows that all occupied levels can be considered to give independent contributions to the polarizability. 

Notably all terms in the sum (\ref{23}) are positive, and the most significant contributions come from energies close to the fermi energy (or chemical potential) $\mu$ where $E_m>\mu$ and $\mu>E_n$ or vice versa such that $f_m p_n \approx 1$ and $\Omn$ is small. On the other hand the rearranged expression (\ref{26}) apparently obtains significant contributions from all levels below the fermi level. But since $\Omn$ can have both signs these contributions thus mostly must cancel.

\section{Induced interactions and perturbation with several particles}
\label{sec7}

With polarizability given by expression (\ref{23}) the results of Sec.~\ref{sec4} are by taking the change of $p_n$ into account,
generalized in a straightforward way. With Eq.~(\ref{25}) one now finds $p_n-p_m=p_n f_m -p_m f_n=2(f_n p_n f_m p_m)^{1/2}\sinh{(\beta\Omega_{mn}/2)}$ since $p_n=(\zeta f_n p_n)^{1/2}e^{-\beta E_n/2}$ and $f_n=(\zeta^{-1} f_n p_n)^{1/2}e^{\beta E_n/2}$. Thus in the  
$(p_m p_n p_k p_l)^{1/2}$ term of Eq.~(\ref{17}) one can replace the $p_i$ ($i=m,n,k,l$) with $f_i p_i$, and with this result (\ref{17}) modifies into
\be
\Delta F=-\psi^2 \Sumn\Sukl f_m p_n f_k p_l \frac{\Mmn\Mkl}{\Omn+\Okl}.
\la{26a}
\ee
Concerning second order perturbation theory the situation may be less obvious. The problem is that one-particle eigenstates can be occupied by other particles. Thus for fermions one may assume that the state of interest is perturbed only via vacant levels. With this modification Eq.~(\ref{18}) for the perturbed energy levels is changed into
\be
\Delta E_{nl}=-\psi^2\sum_{mk}f_m f_k \frac{\Mmn\Mkl}{\Omn+\Okl}.
\la{26b}
\ee

For fermions the grand partition function for the particles in each of the potentials (or atoms), that here are assumed equal, will be
\be
Z=\prod_n(1+\zeta e^{-\beta E_n})
\la{27}
\ee
Thus the total grand partition function with perturbing interaction will be
\bea
Z_t&=&\prod_{n,l}(1+\zeta e^{-\beta E_n}+\zeta e^{-\beta E_l}+\zeta^2 e^{-(\beta E_n+E_l+\Delta E_{nl})}) \nonumber\\
&=&Z^2 (1-\beta\sum_{n,l}p_n p_l \Delta E_{nl}+\cdots)
\la{28}
\eea
with $p_n$ given by (\ref{25}). With free energy $-\beta^{-1}\ln Z_t$ this again leads to the free energy contribution
\be
\Delta F=\sum_{n,l} p_n p_l \Delta E_{nl}
\la{28}
\ee
where now the $p_n$ is given by Eq.~(\ref{25}) and $\Delta E_{nl}$ is given by Eq.~(\ref{26b}). With this one sees that the statistical mechanical result (\ref{26a}) is the same as result (\ref{28}).

\section{Numerical estimate for polarizable atoms}
\label{sec8}

As mentioned earlier the Casimir type energy gives a leading energy contribution to ab initio computations of of molecular energies. This type of contribution, which is a correction to mean field theory, was introduced for classical many-body systems with a perturbing interaction that was assumed weak and of long range where the inverse range $\gamma$ is the perturbing parameter \cite{hemmer64}. For interacting harmonic oscillators it turns out that this correction alone can give the exact answer, so for this case expression (\ref{13}) (before expansion in $\psi$) is exact. 

Clearly, accurate results for the energy corrections to molecular energies will require extensive numerical evaluations on inhomogeneous systems. Thus we here can only give crude estimates. One way is to make an estimate of the van der Waals interaction between a pair of atoms at contact. With static dipolar interaction the $T=0$ result for a pair of harmonic oscillators is after integration with respect to $K$ given by Eq.~(1) of Ref.~\cite{wennerstrom99} as
\be
V(R)=-\frac{3}{4}\hbar\omega_0\frac{\alpha^2 (0)}{R^6}.
\la{40}
\ee
The $\alpha(0)$ can be given a crude estimate from the susceptibility of a polarizable liquid. For low dielectric constant $\varepsilon$ it is approximately given by (in Gaussian units)
\be
\varepsilon-1\approx 4\pi \rho\alpha(0)=4\pi\rho\sigma^3\frac{\alpha(0)}{\sigma^3}
\la{42}
\ee
where $\sigma$ is the (hard core) diameter of an atom. An order of magnitude estimate for the susceptibility may be to put $\varepsilon-1\approx 1.0$ at (dimensionless) fluid density $\rho\sigma^3\approx 0.4$ by which $\alpha(0)/\sigma^3\approx 0.20$. Further one might assume the resonance frequency to give an energy quantum $\hbar \omega_0\approx 1.5$\,eV. With this the van der Waals energy (\ref{40}) at contact $R=\sigma$ would be
\be
V(\sigma)\approx -45\,{\rm meV}.\
\la{43}
\ee
A similar estimate can be obtained for instance on basis of the Lennard-Jones (LJ) interaction $\phi(r)=4\epsilon_{LJ}[(\sigma/R)^{12}-(\sigma/R)^6]$ of Ar (argon). The minimum value of $\phi$ is $\phi_{min}=-\epsilon_{LJ}$. For Ar the critical temperature is  $T_c=151$\,K, and for the LJ fluid $\epsilon_{LJ}/(k_B T_c)\approx 0.8$. Thus with $k_B=1.38\cdot 10^{-23}\,{\rm J/K}=8.63\cdot 10^{-5}$\,eV the attractive van der Waals part of the LJ interaction at contact $R=\sigma$ becomes
\be
-4\epsilon_{LJ}\approx -42\,{\rm meV}.
\la{44}
\ee

\section{Analytic estimate for an electron gas}
\label{sec9}

It will be more satisfying to base the estimate more directly upon the correction to ab initio computations. But for the inhomogeneous fluid of electrons in molecules this does not seem feasible in a simple way. However, we find that an estimate can be made for the homogeneous fluid of electrons (in a neutralizing background). In view of $\gamma$-ordering in terms of the parameter $\gamma$ mentioned above, the electrons without interactions form the reference system, which then is an ideal fermi gas. Since the electrons in the reference system or fluid are free to move the corresponding susceptibility $\varepsilon-1$ will be infinite at zero frequency (as for a metal) in this case; so there is reason to expect a larger correction.

The correction $\Delta F$ to the free energy per unit volume due to the perturbing interaction is given by  Eq.~(5.8) of Ref.~\cite{hoye10}. Now taking spin degeneracy $g=2$ into account (both up and down spins contribute) this becomes
\be
-\beta\Delta F=I=-\frac{1}{2}\frac{1}{(2\pi)^3}\sum_K \int d{\bf k}\,\ln[1-g\hat{S}(K,k)(-\tilde{\psi}(k))].
\la{45}
\ee 
The Fourier transform of the electrostatic Coulomb interaction in SI units is
\be
\tilde\psi(k)=\frac{q^2}{\epsilon_0 k^2}
\la{46}
\ee 
where $q$ is the unit charge of electrons and $\varepsilon_0$ is the permittivity of vacuum. In the uniform case the $\hat{S}(K,k)$ is the structure factor for the ideal gas of free fermions for spins either up or down. (Different spins are uncorrelated in the reference system.) It is given by Eq.~(3.4) of Ref.~\cite{hoye10} or its modified version Eq.~(63) of Ref.~\cite{hoye10a} which is
\be
\hat{S}(K,k)=\frac{\zeta}{(2\pi)^3}\int\frac{\Delta}{K^2+\Delta^2}\frac{X-Y}{(1\pm \zeta X)(1\pm \zeta Y)}\,d{\bf k'}
\la{47}
\ee 
with ($m$ is particle mass)
\bea
\nonumber
&&\Delta=E(k'')-E(k'), \quad E(k)=\frac{(\hbar k)^2}{2m},\\
\la{48}
&&X=e^{-\beta E(k')}, \quad Y=e^{-\beta E(k'')},\\
\nonumber
&&{\bf k''}={\bf k}-{\bf k'}, \quad \zeta=e^{\beta \mu}
\eea 
where $\mu$ is the chemical potential (Fermi energy at $T=0$).

One may expand expression (\ref{45}) to separate out its leading term which is the exchange energy
\be
-\beta\Delta F_{ex}=\frac{g\beta}{2(2\pi)^3}\int\tilde S(0,k)(-\tilde\psi(k))\,d{\bf k}.
\la{49}
\ee 
As given by Eq.~(3.1) in Ref.~\cite{hoye10} (for $\lambda=0$ or $\lambda=\beta$)
\be
\tilde S(0,k)=\frac{1}{\beta}\sum\limits_K \hat S(K,k)=\frac{\zeta}{(2\pi)^3}\int\frac{X}{(1+\zeta X)(1+\zeta Y)}\,d{\bf k'}
\la{50}
\ee 
which is the equal time correlation function. In the limit $\beta\rightarrow\infty$ one finds
\bea
\nonumber
g\tilde S(0,k)&=&\frac{g}{(2\pi)^3}\int\limits_{{k'<k_f}\atop{k''>k_f}}\,d{\bf k'}=\frac{g}{(2\pi)^3}\left[\int\limits_{k'<k_f}\,{\bf k'}-\int\limits_{{k'<k_f}\atop{k''<k_f}}\,d{\bf k'}\right]\\
\la{51}
&=&\rho\left[1-\left(1-\frac{3}{4}u+\frac{1}{16}u^3\right)\right]\quad \mbox{for}\quad 0<u<2,\\
\nonumber
g\tilde S(0,k)&=&\rho \quad \mbox{for}\quad 2<u \quad\mbox{with} \quad u=\frac{k}{k_f}
\eea 
where $k_f$ is the value of $k$ at the Fermi energy, and $\rho$ is the total particle density
\be
\rho=\frac{g}{(2\pi)^3}\int\limits_{k<k_f}\,d{\bf k}=\frac{4\pi g}{3(2\pi)^3}k_f^3.
\la{52}
\ee 
The $\rho$-term in Eq.~(\ref{51}) is a self-energy term ($\delta({\bf r})$-function in ${\bf r}$-space) and should be deleted from Eq.~(\ref{49}) at least with the Coulomb interaction where it will give an infinite contribution. The remaining part of Eq.~(\ref{51}) is the known equal time pair correlation function of fermions at $T=0$. (It is the same as the overlap volume of two spheres each of radius 1 with centers separated by a distance $u$.) Inserting this and expression (\ref{46}) for $\tilde\psi(k)$ gives the known exchange energy per particle ($g=2$)
\be
f_{ex}=\frac{\Delta F_{ex}}{\rho}=-\frac{4\pi }{(2\pi)^3}\frac{3q^2 k_f}{8\varepsilon_0}=-\frac{9(\hbar \omega_p)^2}{32\mu}
\la{53}
\ee 
where $\omega_p$ is the plasma frequency and $\mu$ is the Fermi energy
\be
\omega_p^2=\frac{q^2\rho}{m\varepsilon_0}, \quad \mu=\frac{(\hbar k_f)^2}{2m}.
\la{54}
\ee

The higher order terms of Eq.~(\ref{45}) give the correction to the Hartree-Fock result. In the classical case they give the well known debye-H\"{u}ckel theory of ionic solutions and plasmas where Debye shielding is important. For this situation $g\hat S(0,k)=\beta g\tilde S(0,k)=\beta\rho$ (since then $\beta,\, \rho,\,\zeta X\rightarrow 0$). However, for $T=0$ the correction term turns out to change character. Then the classical term $K=0$ alone gives no contribution. Instead as noted in Ref.~\cite{hoye10} the resulting correlation function reflects the presence of plasma waves. Then the $\Delta$ in the denominator of Eq.~(\ref{47}) was considered negligible and to a first approximation equal to zero. However, to avoid divergence the $\Delta$ can not be neglected here. But  as a simplification the $k$ will be regarded small, by which $\Delta$ can be approximated by
\be
\Delta\approx\frac{-\hbar^2{\bf kk'}}{m}.
\la{55}
\ee
To be able to handle Eq.~(\ref{47}) in an analytic way we have to approximate further. Thus at $T=0$ ($\sum_K\rightarrow(\beta/(2\pi))\int\,dK$)
\be
\frac{1}{2\pi}\int\frac{2\Delta}{K^2+\Delta^2}\,dK=1,
\la{56}
\ee
and as simple approximation Eq.~(\ref{47}) may be written as
\be
\hat S(K,k)=\frac{\Delta^2}{K^2+\Delta^2}\,\frac{2\tilde S(0,k)}{|\Delta|}.
\la{57}
\ee
Used in Eq.~(\ref{50}) (for $T=0$) this gives an identity for $\tilde S(0,k)$ and is thus consistent independent of $\Delta$. But for products of $\hat S(K,k)$ it can only serve as an approximation that may preserve the main features of the energy correction. So some approximation for $\Delta$ related to the integration with respect to ${\bf k'}$ is needed. We choose to take the average of $\Delta^2$ inside the fermi surface. Then
\be
\langle({\bf kk'})^2\rangle=\frac{1}{3}k^2\langle k'^2\rangle=\frac{1}{5}k^2 k_f^2
\la{57a}
\ee
from which with $\mu$ given by Eq.~(\ref{54}) one gets
\bea
\Delta^2=(c_1\mu u)^2 \quad \mbox{with}\quad c_1=\sqrt{\frac{4}{5}}=0.8944,\quad u=\frac{k}{k_f}.
\la{58}
\eea

With $\tilde S(0,k)$ given by Eq.~(\ref{51}) we with approximation (\ref{57}) find (for small $u$)
\be
\hat A(K)=-g\hat S(K,k)(-\tilde\psi(k))=\frac{\Delta^2}{K^2+\Delta^2}\frac{2\rho\frac{3}{4}u}{c_1\mu u}\frac{q^2}{\varepsilon_0 k^2}=\frac{\Delta^2}{K^2+\Delta^2}\frac{\alpha^2}{u^2}
\la{59}
\ee
where
\be
\alpha=c_2\frac{\hbar\omega_p}{\mu}, \quad c_2=\sqrt{\frac{3}{4c_1}}=0.9157.
\la{59a}
\ee
Here it can be remarked that when plasma waves are considered, the $\Delta^2$ in the denominator can be neglected when compared to $K^2$, and one finds $\hat A(K)=-\omega_p^2/\omega^2$ which would require $c_1 c_2=1$ or $c_1=4/3$ in contrast to the value given by Eq.~(\ref{58}). However, in our approximation we will keep the latter as a reasonable value for $\Delta^2$.

Integration of Eq.~(\ref{45}) with respect to $K$ with the linear term taken out further gives
\bea
\nonumber
\tilde f_c(k)&=&\frac{1}{2}\frac{1}{2\pi}\int\limits_{-\infty}^\infty [\ln{(1+\hat A(K))}-\hat A(K)]\,dK\\
&=&\frac{1}{2}\Delta\left[\sqrt{1+\left(\frac{\alpha}{u}\right)^2}-1-\frac{1}{2}\left(\frac{\alpha}{u}\right)^2\right].
\la{60}
\eea

By final integration with respect to ${\bf k}$ the integral will diverge due to our approximation that is limited to small $k$. Thus the integration should be cut, and here the cut will be defined by regarding the low density limit where the ground state energy of plasma oscillations $\hbar\omega_p/2$ will dominate besides the binding energy. The cut with respect to $k$ or $u$ will be made at $u=u_0$ such that each particle gets this ground state energy. For $\alpha$ large the square root term of Eq.~(\ref{60}) times $\Delta$ is then $\Delta(\alpha/ u)=c_1 c_2\hbar\omega_p$. With $(1/(2\pi)^3)\,d{\bf k}=(3/g)\rho u^2\,du$ (with $u=k/k_f$ and $g=2$) this implies
\be
1=\frac{3}{g}\int\limits_0^{u_0}c_1 c_2 u^2\,du=\frac{1}{g}c_1 c_2 u_0^3, \quad u_0=1.3466.\\
\la{61}
\ee

The correction to the free energy, where the linear term is separated out from Eq.~(\ref{45}), now becomes
\be
f_c=\frac{\Delta F-\Delta F_{ex}}{\rho}=\frac{3}{g}\int\limits_0^{u_0}\tilde f_c(k)u^2\,du.
\la{62}
\ee
To integrate it is convenient to introduce a new variable of integration $x=u/\alpha$ with upper limit of integration
\be
x_0=\frac{u_0}{\alpha}.
\la{63}
\ee
With this and relation (\ref{61}) one has
\be
c_1 \mu \alpha^4=c_1 \mu\alpha \frac{g}{c_1 c_2 x_0^3}=\hbar\omega_p\frac{g}{x_0^3},
\la{64}
\ee
so with $\tilde f_c (k)$ given by Eq.~(\ref{60}) we obtain
\be
f_c=\frac{1}{2}\hbar\omega_p\cdot\frac{3}{8}\left[\left(2+\frac{1}{x_0^2}\right)\sqrt{1+x_0^2}-\frac{1}{x_0^3}\ln(x_0+\sqrt{1+x_0^2})-2x_0-\frac{2}{x_0}\right].
\la{65}
\ee
This result is verified by putting $x_0=x$ in expression (\ref{65}) and then differentiate   $x^3 f_c/x_0^3$ with respect to $x$ to obtain $(3/g)\alpha^3 x^2 \tilde f_c(k)$. 

It can be noted that result (\ref{65}) is a valid approximation for any density. For high densities (small $\alpha$) the energy of the electron gas is dominated by the kinetic energy of free fermions and the exchange energy. For low density, however, where $\alpha\rightarrow\infty$ ($x_0\rightarrow 0$)    this is no longer the case. Then the ground state energy of plasma oscillations in addition to the binding energy becomes important. The latter is the exchange energy plus a similar contribution from the $1/x_0$ term at the end of Eq.~(\ref{65}), that may be understood as a contribution from induced correlations.

The above result may have relevance for high-$T_c$ superconductivity where a low density strongly correlated electron gas is expected to play a role. In view of Eq.~(\ref{65}) strong correlations mean strong binding energy plus ground state energy of plasma oscillations while the Fermi energy becomes less significant. Due to the dominance of the binding energy the electrons may prefer to form a regular rigid lattice. This lattice of electrons may adapt to and thus stick to the underlying atomic lattice structure, or it may not do so, depending upon whether net energy is lowered or not. In the first situation the system may be an isolator while in the second case the electron lattice may be free to move relative to the atomic lattice. If so, the system becomes superconducting. However, these speculations, if relevant, will need further investigations.

\section{Numerical estimate for the electron gas}
\label{sec10}

As a numerical example we will consider the energy correction for the electron gas corresponding to one conduction electron per atom in copper (Cu). With atomic weight 63.54 , mass density $8.96\cdot 10^3$\,kg/m$^3$, and Avogadro's number $N_A=6.022\cdot 10^{23}$\,mol$^{-1}$ the particle density becomes
\be
\rho=8.5\cdot 10^{28}\,{\rm m}^{-3}
\la{66}
\ee
Further with electron mass $m=9.11\cdot10^{-31}$\,kg, electron charge $q=1.602\cdot10^{-19}$\,As, $\varepsilon_0=8.854\cdot10^{-12}$\,(As/(Vm)), and $\hbar=1.054\cdot10^{-34}$\,Js one finds (1\,eV$=1.602\cdot 10^{-19}$\,J)
\be
\hbar\omega_p=10.82\,{\rm eV}, \quad \mu=7.04\,{\rm eV}.
\la{67}
\ee
Thus with Eqs.~(\ref{63}), (\ref{61}), and (\ref{59a}) one gets
\be
x_0=\frac{u_0}{c_2}\frac{\mu}{\hbar\omega_p}=0.957.
\la{68}
\ee
When this is inserted in Eq.~(\ref{65}) the correction or exchange-correlation energy becomes
\be
f_c\approx-1.4\,{\rm eV}.
\la{69}
\ee
As a comparison the average particle energy of the reference system free electron gas is $(3/5)\mu=4.22$\,eV while the exchange energy (\ref{53}) is $f_{ex}=-4.68$\,eV. Thus for the uniform electron gas the estimated magnitude of the correction beyond the exchange energy is quite significant for the density of conduction electrons in metals. In view of result (\ref{65}) this significance will increase for lower densities of the electrons (at $T=0$). 

It here can be noted that when comparing the results of Ref.~\cite{hoye10} with other approaches one finds that the Casimir energy, at least when electrostatic interactions are used, is equivalent to the results of the random phase approximation (RPA) for quantized many-body systems \cite{pines52, nguyen10}. With the RPA it is expected that the energy functional should be optimized or made self-consistent with respect to its RPA part too. This results in a formidable computational task which limits the method to simple and small systems \cite{nguyen10, lein00}. However, from the viewpoint of Casimir energy used in this work there should be no need for such optimization of the RPA part since from the statistical mechanical derivation and $\gamma$-ordering \cite{hemmer64} this part is the leading correction to HF and DFT computations. Its influence upon wave functions will be similar. At optimum a change in wave functions will to linear order in this change have no influence upon the energy; it will be quadratic and by that of higher order. Thus it can be neglected, and the formidable computational task just mentioned can be avoided. Likewise no such optimization were needed to obtain the Casimir energy where possible non-linear parts of the dielectric constant are neglected. This is clearly the case for parallell plates filled with electrons where change in charge density at the surfaces due to electron interactions were neglected \cite{hoye08}. Thus in this respect similar changes in molecules can be neglected.

The RPA is the quantized version of the Debye-H\"{u}ckel theory for ionic systems. As pointed out by Lein et al.~the RPA approximation gives too low energy \cite{lein00}. (The situation is similar for the classical Debye-H\"{u}ckel theory.) 
This they correct by modifications of the correlations at short range. Related modifications in view of the statistical mechanical picture are indicated at the end of Sec.~5 of Ref.~\cite{hoye10}. Especially we here have in mind the simple properties of the direct correlation function that have been experienced for classical systems. These properties are expected to carry over to the "polymer" path integral of quantized systems as has been demonstrated for hard core classical polymers \cite{hoye04}. Then the hard core condition defines the direct correlation function or an effective interaction inside the hard core diameter via the Ornstein-Zernike integral equation. The perturbed correlation function of electrons (for the mixture of spins up and down) should also fulfill a hard core condition for small r (exact for $r=0$). The $f_{xc}(q,\omega)$ function (or the corresponding local field factor $G(q,\omega)$)) of Ref.~\cite{lein00} serves a similar purpose to obtain an effective interaction, and various approximations to it have been considered. As far as we can see they do not include the use of a hard core or an effective hard core condition which also can be applied in the non-uniform case.

In Ref.~\cite{lein00} the various corrections to the RPA were compared with "exact" results. One such case is for $r_s=1$ with
\begin{equation}
\frac{4\pi}{3}(r_s a_0)^3=\frac{1}{\rho}
\label{70}
\end{equation}
where  $a_0=4\pi\varepsilon_0 \hbar^2/(mq^2)=5.29\cdot10^{-11}$\,m is the Bohr radius. This gives particle density
\begin{equation}
\rho=1.61\cdot10^{30}\,\rm{m}^{-3},
\label{71}
\end{equation}
for which we find
\begin{equation}
\hbar\omega_p=47.1\,\rm{eV}, \quad \mu=50.1\,\rm{eV}, \quad x_0=1.564
\label{72}
\end{equation}
with correction or exchange-correlation energy
\begin{equation}
f_c\approx-2.4\,\rm{eV}.
\label{73}
\end{equation}
Various "exact" results referred to in Ref.~\cite{lein00} are -1.63\,eV, -1.62\,eV, -1.53\,eV, and -1.64\, eV. In view of the approximations made, result (\ref{73}) is reasonable. 

However, the approximate simple expression (\ref{60}) can be a basis for more accurate evaluations. In this respect we have made a comparison with Fig.~2 of Ref.~\cite{lein00} where the energy distribution $\varepsilon_c(q)=3f_c(k) u^2$ ($q=k$) is plotted as function of $q/(2k_f)=u/2$ for the case $r_s=4$. For small $u\rightarrow 0$ our $f_c$ follows both the RPA and "exact" curves, while for larger $u$ it becomes more negative as reflected in result (\ref{73}). Then to obtain the RPA curve one in approximation (\ref{57}) can replace the $\tilde S(0,k)$ with an effective quantity (as a substitute for exact evaluations). Further to obtain the "exact" curve from the RPA curve the interaction (\ref{46}) is modified for large $k$. This modification is a small $r$ contribution to the direct correlation function (which indirectly defines the $f_{xc}(k,\omega)$ mentioned above). So multiplying the (\ref{46}) with the simple factor $1/(1+ck^2)^2$ ($c=$const.) already gives a good approximation to the exact curve. The challenge and numerical task will be to predict quantities like the constant  $c$ on basis of the hard core condition mentioned above. The predictive power of the hard core condition will be especially useful for the non-uniform case of electrons forming molecules where "exact" results for comparisons are less available.

\section{Summary}
\label{sec11}

We have evaluated the induced van der Waals interaction between two point particles (e.g.~two atoms) by a statistical mechanical method. Comparing with standard quantum mechanical perturbation theory to second order the results are the same. This seems obvious from the physical side, but mathematically this is not so. The equivalence is restricted to short separations where retardation effects can be neglected. 

The statistical method can also be applied when retardation effects are present, and the resulting interaction is the Casimir-Polder one. The equivalence shows that the Casmir interaction as obtained by the latter method is an effect relevant for quantum mechanical evaluations of many-body systems like many-electron molecular systems. The leading term in the expansion of Eq.~(\ref{13}) corresponds to second order perturbation theory. In addition there are higher order terms, and time dependent interactions like the electromagnetic one can be included. Thus in recent works we added the corresponding perturbation to ab initio calculations of systems like the electron clouds of molecules \cite{hoye10,hoye10a}. This perturbation gives the leading energy correction to the well-known Hartree-Fock and density functional theory \cite{kohnsham65}. In this way induced energies due to non-local correlations are accounted for, and shielding of electric charges, which is standard Debye-shielding in the classical case, is embedded when contributions corresponding to all terms of expression (\ref{13}) are included. 

To see the magnitude of this perturbation we have estimated the inter-atomic induced energy for a pair of atoms at close contact. On basis of their van der Waals  interaction this energy was found to be of the order of $-40$\,meV. Further we have analyzed the equations for the uniform electron gas and obtained explicit approximations (for $T=0$). Especially the results are valid for the electron gas at low density where plasma oscillations become more significant and may have relevance for high-$T_c$ superconductivity. For the electron density of conduction electrons in metals the approximation obtained gave an energy correction of roughly $-1$\,eV.

For some higher density of electrons comparison is made with an "exact" result. It is noted that the Casimir energy is closely associated with the RPA for which the statistical mechanical approach will indicate new ways to perform modifications
to obtain accurate results.
\ \\
\ \\
{\large\bf Acknowledgement}\\
\ \\
We are indepted to Professor Iver Brevik for useful comments to this work in view of his knowledge and insight into the various aspects and problems related to the Casimir effect.


\end{document}